# Embedded Development Boards for Edge-AI: A Comprehensive Report

Hamza Ali Imran
Department of Computing
School of Electrical Engineering & Computer Science,
National University of Sciences and Technology (NUST), Islamabad, Pakistan
himran.mscs18seecs@seecs.ed.pk

Usama Mujahid
Department of embedded systems,
RwR Private Limited, I-9 /3, Islamabad, Pakistan
umujahid363@gmail.com

Saad Wazir
Department of Computing
School of Electrical Engineering & Computer Science,
National University of Sciences and Technology (NUST), Islamabad, Pakistan
swazir.mscs18seecs@seecs.edu.pk

Usama Latif
Operations Engineer
VAS, Apollo Telecom,
Islamabad, Pakistan
usamalatif417@gmail.com

Kiran Mehmood
Department of Computing
School of Electrical Engineering & Computer Science,
National University of Sciences and Technology (NUST), Islamabad, Pakistan
kmehmood.mscs19seecs@seecs.edu.pk

*Abstract*—The use of Deep Learning and Machine Learning is becoming pervasive day by day which is opening doors to new opportunities in every aspect of technology. Its application Ranges from Health-care to Self-driving Cars, Home Automation to Smart-agriculture and Industry 4.0. Traditionally the majority of the processing for IoT applications is being done on a central cloud but that has its issues; which include latency, security, bandwidth, and privacy, etc. It is estimated that there will be around 20 Million IoT devices by 2020 which will increase problems with sending data to the cloud and doing the processing there. A new trend of processing the data on the edge of the network is emerging. The idea is to do processing as near the point of data production as possible. Doing processing on the nodes generating the data is called Edge Computing and doing processing on a layer between the cloud and the point of data production is called Fog computing. There are no standard definitions for any of these, hence they are usually used interchangeably. In this paper, we have reviewed the development boards available for running Artificial Intelligence algorithms on the Edge.

*Keywords—Internet of Things (IoT), Edge Computing (EC), Artificial Intelligence (AI), AI on the Edge, MPSoc (Multi-processor system on chip), APU (Application Processing Unit), RPU (Real-time Processing Unit), GPU (Graphics Processing Unit), Neural Network Processor (KPU)*

I. INTRODUCTION

Artificial Intelligence (AI), Machine Learning (ML) and Deep Learning (DL) are playing an important role in many applications nowadays. Self-driven cars are not fiction anymore. Computers have surpassed human's capability to classifying images in ImageNet Challenge [1]. Since the invention of AlexNet (Convolution Neural Network) in 2012 Computer Vision has entered a new era. Deep Neural networks are state of the art techniques for applications including Computer Vision, Natural Language Processing, Voice Recognition, etc. They are making computers perform tasks that were once considered impossible. AI is becoming pervasive day by day; contributing towards Smart-cities, newer industrial revolution called Industry 4.0, Smart Health-care systems, Agriculture, Education, and the list goes on. Internet connectivity is becoming a part of every household device. There is explosive growth in internet-connected devices and it is expected that by 2020 there will be 20 Million IoT devices around the globe [2].

Traditionally processing is being done in the Cloud. IoT devices majorly played a role in data picking and actuating; decision making was being done on Central Cloud. This has its problems including privacy, the demand for large bandwidth, latency, etc. Recently, newer concepts of Edge and Fog Computing are being introduced and the target of these technologies is to maximize processing near the point of data generation. Edge computing loosely means to do processing on the device or node and Fog means to do processing on the edge of the network which means a layer that is in between the cloud and IoT device. There are no standard definitions of both of these technologies hence they are mostly used interchangeably but Edge Computing is considered to come under the umbrella of Fog Computing. Embedded Systems inside the IoT devices have their limitations. They are very limited in compute and storage resources.

Moreover, most of the devices are battery operated and are to stay ON for long periods. The more we are near to cloud the better the amount of computing and storage resources we get but this results in poor latency which in case of applications having hard timing conditions like self-driving cars is not suitable. Figure 1 does a comparison between the use of Cloud computing and Edge or Fog Computing.

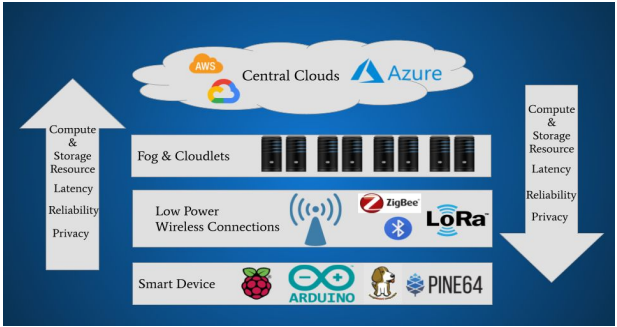

[Fig. 1]: Comparison of Cloud versus Edge and Fog Computing

AI models are always trained on the Cloud as the training process requires a lot of resources and takes a lot of time. Training means to find the value of parameters in the modules; which in case of current time state of the art Neural Networks ranges from a few hundred to millions. The trained model is then expected to run on an embedded device.

Work is being done on both sides; software and hardware, to make the embedded device run Deep Neural Networks efficiently. On the software side, a technique called Quantizing [3] is being used. It means to store weights in compact formats such as integers or binary numbers instead of floating-point numbers to reduce the size of the model. Moreover, quantization is also targeted towards replacing arithmetic operations with floating-point operations which results in less energy consumption. On the hardware side, new processors are being designed to run AI models efficiently. In this paper, we have reviewed the embedded development boards that are specially designed to run AI models.

## II. TYPES OF AVAILABLE HARDWARE

DARPA (Defense advanced projects agency) categorizes the progress of AI in three waves. The first wave is the beginning of the revolution where the machine follows the set of rules. This includes the traditional way of programming. The second wave is about statistical learning where deep learning models and the learning data is provided to the machine and machine programs itself. This was a revolution and resulted in wonders like self-driving cars. The third wave is a futuristic idea in which the machines will create their own models and explain how the world works.

We are currently going through the second wave and edging closer to the third one. This implies we have to continuously evolve our AI systems to run the algorithms efficiently. There are different types of hardware available for AI. It consists of mainly CPU, GPU, FPGA, and ASICs. Every hardware has its pros and cons and we have to choose a suitable trade-off. Figure 2 shows that the CPUs are the most flexible and the least efficient in running the ML while the ASICs like TPU is the most efficient but the least flexible among all the options.

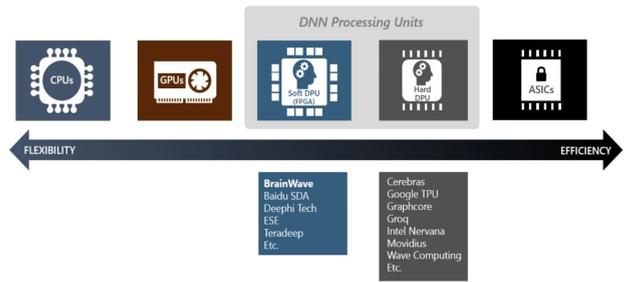

[15, Figure 2] Different alternatives of Deep neural network

Edge AI is the need of the hour and every Big name like NVIDIA, Xilinx, Intel etc has launched their boards to capture the market. We have reviewed the most well-known boards from these companies. We have divided these boards into different categories based on the type of hardware they use for ML acceleration. It is shown in table 1

| Type | Boards |
|---|---|
| GPU | Beaglebone AI |
| | Nvidia Jetson nano |
| ASIC | Google edge |
| | Sophon edge |
| FPGA & GPU | Ultra96 |
| ASIC & Microcontroller-based | Sipeed Maxduino |

[ Tab. 1] Categories of Boards

## III. OVERVIEW OF BOARDS

This section has a comprehensive review of the technical specification of development boards for AI on the Edge.

### ASIC Accelerated Boards

*A. Google Edge TPU*

Google Edge TPU is the part of google IoT. They are specifically designed to run inferences at the Edge with the help of machine learning models that are pre-trained at the cloud. The Coral Development Board which is integrated with the Google Edge TPU which is ASIC (Application-Specific Integrated Circuit) design to support on-device ML (Machine Learning). Coral Development Board is an SBC ( Single Board Computer) for High-speed Machine learning inferencing and has Wireless capabilities. It has a removable SoM (System-on-Module). This board runs an OS which is a variant of Debian Linux called Mendel.

The Edge TPU coprocessor is capable of performing 4 trillion Operations Per Second (TOPS), using just 0.5 watts for each TOPS. It can support both C++ and Python programming languages. It uses Mendel Development Tool (MDT) which is a command-line tool to perform tasks with connected Mendel devices. It supports Google TensorFlow Lite and AutoML Vision Edge. You can only use Python API and C++ API to perform inferences with TensorFlow Lite models. Python API can be used by importing the edgetpu module and C++ API can be used by including

edgetpu.h header file. This Dev board is mainly used for image classification and object detection but it can be used for other purposes. Good documentation and support make it easier to work with this Dev Board.

[5] implemented a real-time recognition for image classification and get instant results, this dev board has a great potential for performing real-time ML calculations. [6] Implemented object detection demo from the video and image classification using a camera module. The following table (table 2) has its hardware specifications.

| Processing unit/s | |
|---|---|
| CPU | NXP i.MX 8M SoC (quad Cortex-A53, Cortex-M4F) |
| AI accelerator/s | |
| GPU | Integrated GC7000 Lite Graphics |
| ML accelerator | Google Edge TPU coprocessor |
| Memory | |
| RAM | 1 GB LPDDR |
| Flash memory | 8 GB eMMC + MicroSD slot |
| GPIO and Interfaces | |
| Interfaces | Wi-Fi 2x2 MIMO (802.11b/g/n/ac 2.4/5GHz) and Bluetooth 4.2 , HDMI 2.0a (full size); 39-pin FFC connector for MIPI-DSI display (4-lane) , Gigabit Ethernet port |
| GPIO | 3.3V power rail; 40 - 255 ohms programmable impedance; ~82 mA max current |
| Power Requirements | |
| Voltage and Current Requirement | 5V DC (USB Type-C) |

[4, Tab. 2] Hardware specifications of Coral Dev Board

*B. Sophon Edge*

Sophon Edge is a development board released by BITMAIN which is leading vendors of Bitcoin mining computers and chips. This board brings powerful deep learning tools to a variety of applications with quick prototype development. It comes with a powerful BM1880 chip which can implement DNN/LSTM/RNN/CNN models and operations with ease. It has a TPU (Tensor Processing Unit), 1.5 GHz dual-core a53 and RISC CPU based on RISC-V architecture. This board is built in accordance with Linaro 96boards consumer edition. One of the main advantages of this board is its ultra low power architecture that helps in running complex algorithms efficiently. Sophon Edge is the first ASIC based 96board. It uses Edge TPU to act as the main AI accelerator. BM1880 can act both as a main processor or as a co-processor for deep learning.

BITMAIN claims [10] that it can receive data from other hosts to perform inference and return the results to the host. Its block diagram (Figure 3) and hardware details (Table 5) are given below.

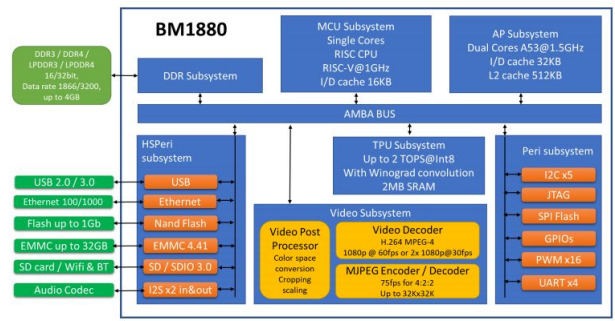

[10, Fig. 3] Block Diagram of Sophon Edge

| Processing unit/s | |
|---|---|
| Processor 1 | 2x Cortex A-53 @ 1.5GHz |
| Processor 2 | RISC V @ 750 MHZ |
| AI accelerator/s | |
| TPU | Up to 2 TOPS (Tera operations per second) with Winograd convolution |
| Memory | |
| RAM | 1 GB LPDDR |
| EMMC | 8 GB |
| GPIO and Interfaces | |
| Interfaces | Gigabit Ethernet,Wifii, Bluetooth, USB camera (UVC), UART, JTAG, 3x USB 3.0, I2S |
| GPIO | 40 pin low-speed expansion header |
| Power Requirements | |
| Voltage and Current Requirement | 12V @ 2A |

[11, Tab. 5] Hardware specifications of Sophon Edge

The machine learning frameworks support by it include TensorFlow, Pytorch, ONNX, Caffe, MXNet. BITMAIN recommends [11] this board for applications involving object recognition and detection, facial recognition, voiceprint recognition, License plate recognition. BITMAIN has launched BMNNSDK (BITMAIN Neural Network Software development kit). This is a recommended environment to implement machine learning projects. This environment is created to maximize inference throughput and efficiency. It contains BMNET and BMRunTime. BMNET is a deep neural network compiler designed for edge TPU processor. It converts algorithms like CNN to TPU instruction. BMRunTime is actually a powerful library created to provide different software interfaces through a set of APIs. It obscures hardware details and makes programming easy.

**ASIC & Microcontroller Based Boards**

## C. Sipeed Maixduino

This board has a form factor of Arduino Uno. It is like an Arduino for Deep learning projects. Has MAIX SoC and ESP32 on it. Sipeed claims that MAIX is the world's first RISC-V 64 bit based AI module which has a 1x1 inch form factor. MAIX is designed for AI applications it has two RISC-V 64 CPU cores, a dedicated audio processing Core (APU) and dedicated KPU (Neural Network Processing Unit) which is optimized to run Convolutional Neural Networks. Convolutional Neural Networks are state of the art Deep Neural Networks for Image recognition. MAIX also has an FFT unit for performing Fast Fourier transforms making is suitable for signal processing as well. Although with these it also has many other peripherals support as well. Figure 4 shows the block diagram of MAIX SoC.

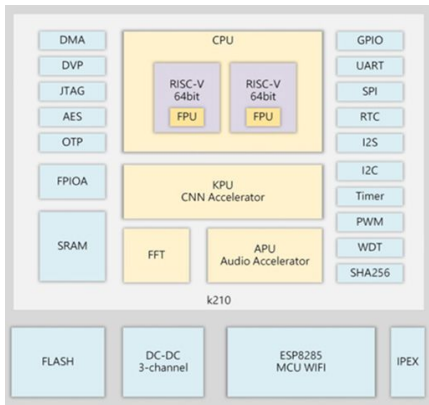

[8, Fig. 4] Block Diagram of MAIX SoC

Following table (table 3) has the hardware specification of Maixduino Board

| System On Chip/s | |
|---|---|
| AI acceleration | Sipeed MAIX |
| Microcontroller | ESP32 |
| **Interfaces** | |
| DVP Camera Interface | 249 0.5mm FPC connector |
| LCD Connector | 8bit MCU LCD 24P 0.5mm FPC connector |
| Wireless Standard | 802.11 b/g/n |
| Onboard MEMS microphone | MSM261S4030H0 is an omnidirectional, Bottom-ported, I2S digital output MEMS Microphone |
| **Power Requirements** | |
| Voltage and Current Requirement | 4.8V ~ 5.2V & I >600mA |

[7, Tab. 3] Hardware specifications of Maixduino Board

This board works with MaixPy IDE, Arduino IDE, OpenMV IDE, and PlatformIO IDE Integrated development environments. Deep learning Frameworks Support includes Tiny-Yolo, Mobilenet, and TensorFlow Lite.

## FPGA & GPU Accelerated Boards

## D. Ultra96

Ultra96 is a Xilinx development board based on Linaro 96boards consumer edition. It is based on Arm-based Xilinx Zynq UltraScale + MPSoC architecture. This board is designed for Artificial intelligence, Machine learning, and IoT applications. It uses the processing power of GPUs and FPGA based accelerators to implement the AI efficiently.

Petalinux and Pynq are recommended and the most widely used Linux distribution for the Ultra96 board. Out of the box images for these distributions are also available on their website. Pynq is more suitable for machine learning applications. Table 4 shows the hardware specifications of the Ultra96 board. and block diagram taken from Ultra96's manual is given as figure 5

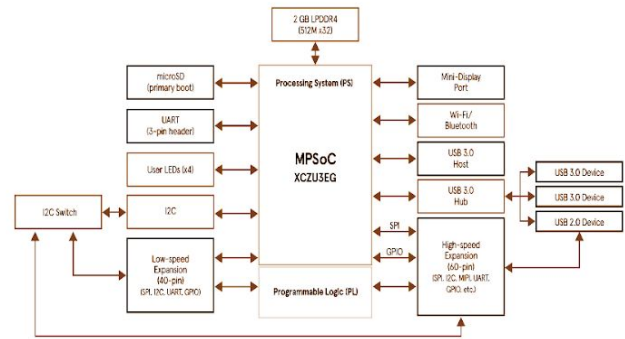

[13, Fig. 5] : Block Diagram of Ultra96 Block diagram

| Processing unit/s | |
|---|---|
| Application processing unit (APU) | 4x Cortex A-53 @ 1.5 GHz |
| Real-time processing unit (RPU) | 2x Cortex R5 |
| **AI accelerator/s** | |
| FPGA | ZU3EG A484 FPGA |
| GPU | Mali 400 MP2 |
| **Memory** | |
| RAM | 2 GB LPDDR |
| EMMC | 8 GB |
| **GPIO and Interfaces** | |
| Interfaces | Bluetooth 4.2, Wifi, 1x USB 3.0 upstream, 2x USB 3.0 downstream, mini Displayport, UART, JTAG |
| GPIO | 60 High speed and 40 low-speed expansion headers |
| **Power Requirements** | |
| Voltage and Current Requirement | 8V ~ 18V @ 3A |

[Tab. 4] Hardware specifications of Ultra96

Ultra96 supports both open source and Xilinx proprietary softwares. Supported machine learning frameworks are

Caffe, MxNet and TensorFlow Lite. This board has been widely used in AI and ML projects. In a project named "Caffein-AI-tor" by Dnhkng [8]. He implemented double deep learning CNNs with face and emotion recognition on the Ultra96 board. This project recognizes the face and emotion of the person and predicts which type of coffee the person wants at that time. Random Forest algorithm is used in the end to predict the choice, keeping in view of past history. In another project Speech recognition at the edge is implemented by Cvoinea [9]. In this project DeepSpeech TF (Tensorflow) pre-trained model is used to implement DNN (Deep Neural Network). TF model is implemented on the A53 core with accelerators running on the FPGA.

## GPU Accelerated Boards

### E. Nvidia Jetson Nano

Neural Networks need to do calculations that are embarrassingly parallel in nature this makes GPUs (Graphics Processing Units) a good choice for running them. Nvidia has recently released Jetson Nano which is a development board that has 128 Maxwell's GPU cores in it. It requires 5 to 10 Watts of energy depending on the task. Nvidia has created a Ubuntu 18.04 LTS Image for their Jetson Nano Developer Kit which has major deep learning frameworks and tools already in it. Jetson Nano supports all major deep learning frameworks and tools famous ones include TensorFlow, pyTorch, Caffe/Caffe2, Keras, MXNet.Since its release, Jetson is being used by many for AI on the Edge projects. A very famous project which is made open source is "Jetbot" it is a robot that takes input through a camera then runs it through a neural network and does obstacle avoiding accordingly. Figure 6 given below shows Jetson Nano Kit and table 5 shows its hardware details

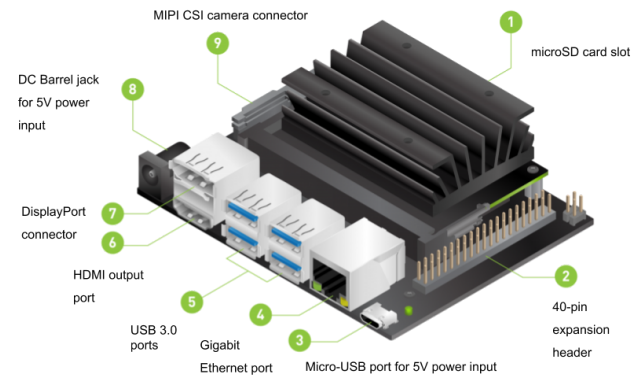

[Fig. 6] Block Diagram of Nvidia Jetson Nano

| Processing unit/s | |
|---|---|
| The central processing unit (CPU) | Quad-core ARM Cortex®-A57 |
| AI accelerator/s | |
| GPU | 128 Nvidia Maxwell Cores |
| Memory | |
| RAM | 4 GB 64-bit LPDDR4 |
| EMMC | 16 GB |
| GPIO and Interfaces | |
| Interfaces | Gigabit Ethernet, HDMI 2.0 or DP1.2 \| EDP 1.4 \| DSI (1 x2) 2 simultaneous |
| GPIO | 1x SDIO / 2x SPI / 4x I2C / 2x I2S / GPIOs -> I2C, I2S |
| Power Requirements | |
| Voltage and Current Requirement | 5V @ 2A |

[12, Tab. 5] Hardware specifications of Jetson Nano Development Kit

### F. BeagleBone AI

BeagleBone AI is a very new, sophisticated, high-end single board computer. It is designed to reduce the gap between normal SBCs and powerful industrial computers. It is targeted for machine learning and computer vision applications. It has a Texas Instruments AM5729 SoC. It is very versatile in processors and accelerators and can support a range of applications. It supports all the famous machine learning frameworks like TensorFlow, Caffee, MXNET, etc. Figure 7 shows the block diagram and table 6 shows its hardware description. It has the capability to run the ML algorithm efficiently by utilizing its IPU and GPU accelerators. It has an EVE subsystem that is specialized to run the computer vision application. BeagleBone recommends using Cloud9 online IDE for programming purposes.

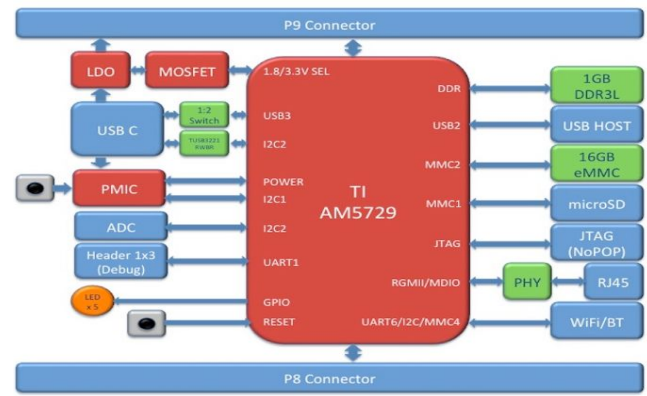

[14, Fig. 7] Block Diagram of BeagleBone AI

| Processing unit/s | |
|---|---|
| MPU(micro-processing unit) subsystems | Dual Cortex A-15 MPU @ 1.5GHZ |
| DSP(digital signal processor) subsystems | Two C66x VLIW Co-processors |
| EVE(Embedded vision engine) subsystems | Four EVE analytics processor |
| PRU (programmable real-time unit) subsystems | Dual 32-bit RISC cores |
| IPU (intelligence processing unit) | Dual Cortex M4 cores |
| **AI accelerator/s** | |
| 3D GPU | POWERVR SGX544 |
| 2D graphics accelerator | Vivante GC320 |
| **Memory** | |
| RAM | 1 GB |
| EMMC | 16 GB |
| **GPIO and Interfaces** | |
| Interfaces | Gigabit Ethernet, wifi 802.11ac 2.4/5G HZ, 4x UART, 2x I2C, 2x SPI, 1x USB Type-C port, 1x USB Type-A host |
| GPIO | 2x 46 pin expansion headers |
| **Power Requirements** | |
| Voltage and Current Requirement | 5V @ 3A |

[15, Tab. 6] Hardware specifications of BeagleBone AI

## IV. Pros and Cons of Boards

Table 7 shows the pros and cons of all the boards we have considered for review in this paper.

| Board name | Pros | Cons |
|---|---|---|
| Google Edge TPU | It has integrated support for TensorFlow and other Google APIs. It also supports removable SoM to scale the production | Supports only TensorFlow lite which is the lighter version of TensorFlow, |
| Sipeed Maixduino | Consume very less power relatively and cost-effective | Not much documentation available |
| Nvidia Jetson Nano | Uses Nvidia's GPU which are traditionally used for deep learning hence support models of all famous frameworks | No significant problem. |
| Ultra96 | Uses both FPGA and GPU to accelerate the ML | No performance issue. A bit expensive |
| Sophon Edge | Uses TPU and gives a superior performance with TensorFlow also a power-efficient device | Limited options in the software development environment |
| BeagleBone AI | Contains a variety of processing units dedicated to specific tasks | No significant issue to this date |

[Tab. 7] Pros and Cons of all the Boards

## V. Conclusion

The use of AI is becoming common for IoT devices which results in many exciting applications. Neural Networks require a lot of computing resources for that purpose the majority of the processing is being done on the Cloud which has many potential problems including privacy, latency, etc. Processors for running AI algorithms efficiently are appearing in the market. We have reviewed some of the most renowned boards that are available in the market. The use of such processors can reduce the traffic to the cloud and can help in running AI on the Edge. They are also acting as a pioneer in spreading the AI in daily life.